\documentclass[twocolumn,superscriptaddress,aps,prl]{revtex4-1}

\usepackage{tabularx}
\usepackage{amsmath}
\usepackage{amssymb}
\usepackage{bm}
\usepackage{graphicx}
\usepackage{psfrag}
\usepackage{url}
\usepackage{color}

\begin{document}

\title{Enhanced motility of a microswimmer in rigid and elastic confinement} 
\author{\firstname{Rodrigo} \surname{Ledesma-Aguilar}}
\email{r.ledesmaaguilar1@physics.ox.ac.uk}
\author{\firstname{Julia M.} \surname{Yeomans}}
\affiliation{The Rudolf Peierls Centre for Theoretical Physics, University of Oxford, 1 Keble Road, Oxford OX1 3NP, United Kingdom}
\date{\today}

\begin{abstract}
We analyse the effect of confining rigid and elastic boundaries on the motility of a model dipolar microswimmer. 
Flexible boundaries are deformed by the velocity field of the swimmer in such a way that the motility of both extensile and contractile swimmers is enhanced. 
The magnitude of the increase in swimming velocity is controlled by the ratio of the swimmer-advection and elastic timescales, and the dipole moment of the swimmer.  
We explain our results by considering swimming between inclined rigid boundaries. 
\end{abstract}

\maketitle

{\it Introduction:--} 
Confinement in rigid and elastic environments is a key concept affecting fluid transport and locomotion in 
microscopic systems, ranging from molecular motors to single and multicellular self-propelled 
organisms~\cite{Lauga-RepProgPhys-2009}. 
The low-Reynolds number world of micro-swimmers is often crowded by passive and active, permeable and impermeable,
boundaries, such as viscoelastic gels, microtubules or cell walls. While these
can act as barriers or defence mechanisms against microorganisms, it has been 
suggested that they can be exploited by the swimmers to enhance their motility~\cite{Lauga-RepProgPhys-2009}. 
Not surprisingly, the locomotion of pathogens close to flexible and rigid surfaces
can play a key role in their success or failure as infectious agents.  
Motile pathogenic bacteria and parasites move through the extracellular matrix of the host to invade 
the vascular system~\cite{Moriarty-PLOS-2008} and, in the case of neurodegenerative infections,
even cross the blood-brain barrier~\cite{Pulzova-FEMS-2009}.
For example, pathogenic sphirochaetes responsible for syphilis~\cite{Peeling-JPathol-2006}, leptospirosis~\cite{McBride-CurrOpinInfectDis-2005} 
and Lyme disease~\cite{Charon-AnnRevMicrobiol-2012}, have been reported to swim in microvasculature channels~\cite{Moriarty-PLOS-2008}, renal tubules~\cite{McBride-CurrOpinInfectDis-2005}, 
and to invade the intracellular junctions between endothelial cells~\cite{Thomas-PNAS-1988}. 
Defensive mechanisms against infections also involve close interaction between complex elastic surfaces
and pathogens.  Flexible-surfaced leukocytes prey on motile microorganisms~\cite{Nourshargh-NatRevMolCellBio-2010}, 
such as {\it E. coli} and {\it P. aeruginosa}, and the viscoelastic gel lining the stomach acts as a barrier against 
{\it H. pylori}~\cite{Montecucco-NatRevMolCell-2001}, a bacterium linked to chronic gastritis and stomach cancer.
Understanding the way in which microswimmers behave in confinement can also lead to novel applications in microfluidics and biotechnology, 
ranging from bacterial rectification~\cite{Drescher-PNAS-2011,Kantsler-PNAS-2012} to controlled steering of artificial micro-robots 
~\cite{Sudo-JIntelMatSystStr-2006,Martel-IntJRobotRes-2009}. 

\begin{figure}[b!]
\psfrag{xlabel}[c][c][1]{$\tau_{\rm s}/\tau_{\rm f}$}
\includegraphics[width=0.44\textwidth]{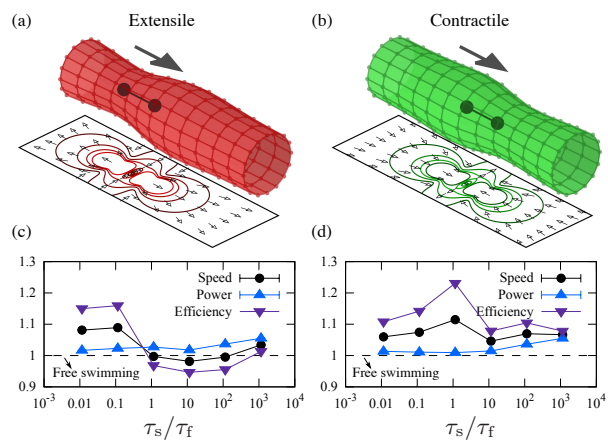}
\caption{\label{fig:3DAll} 
(Colour online) (a) and (b) Simulations of dipolar extensile (pusher) and contractile (puller) swimmers passing through
an elastic tube. The deformation of the channel, protruding on average towards pushers and away from pullers, 
is set by the competition between the swimmer velocity field 
(projected planes) and the elastic response of the boundary.  (c) and (d) The increased friction  
due to the solid boundaries leads to a change in the swimmer velocity, the power consumed and the swimming
efficiency, relative to the free-swimming case.
The amplitude of the response in motility depends on the
ratio between swimmer and elastic timescales, $\tau_{\rm s} /\tau_{\rm f}$. }
\end{figure}

Motivated by these observations, in this Letter we formulate the following questions: what is the effect of the activity of a microscopic swimmer on a bounding
elastic surface and, what is the effect of the interaction with the surface on the motility of the swimmer?
While previous efforts have mainly focused on swimming strategies that exploit rigid and elastic surfaces to overcome the scallop
theorem~\cite{Trouilloud-PRL-2008,Crowdy-JFluidMech-2011,Spagniole-JFM-2012}, the effect of confinement by rigid 
and flexible boundaries on the motility of microswimmers remains largely unexplored~\cite{Katz-JFM-1974,Fauci-Bull-1995,Zhu-JFM-2012}.  
By providing a microscopic footing, single-particle interactions can be used to build up coherent long-wavelength 
hydrodynamic models of active matter subject to viscoelastic heterogeneity~\cite{Leshansky-PRE-2009,Fu-EPL-2010} and
confinement~\cite{Leoni-PRL-2010,Brotto-PRL-2013,Furthauer-PRL-2013,Wioland-PRL-2013} to model, {\it e.g.,} peristaltic pumping and cytoplasmic 
streaming~\cite{Woodhouse-PRL-2012,Ravnik-PRL-2013}.

We begin our discussion by presenting {3D hydrodynamic simulations} of a single microswimmer moving along the centre line of an elastic 
tube~[Figs.~\ref{fig:3DAll}(a) and (b)].
{The surface of the tube is a rectangular mesh of points, coupled by stretching and bending elastic springs
which, in equilibrium, form a cylinder.
By choosing the mesh size and elastic coupling between points on the surface we can model both impermeable and permeable tubes, as well
as sets of uncoupled or cross-linked filaments which mimic gel-like environments [see Supplementary Information for more details of the model]. 
}
The swimmer is composed of two beads {of variable hydrodynamic friction coefficient} coupled by 
an elastic spring whose rest length oscillates in time according to a prescribed swimming stroke~\cite{Avron-NJPhys-2005}. 
The resulting force-free swimmer acts on the fluid as a force dipole of strength $p$ and has an advection
timescale $\tau_s$ that depends on the swimming stroke.
Such generic dipolar swimmers~\cite{Lauga-RepProgPhys-2009} can be `pushers' or `pullers', depending on their flow pattern~[projections in Figs.~\ref{fig:3DAll}(a) and (b)]. 
For pushers this is extensile (fluid is pushed out from the ends of the swimmer and drawn in to the sides) 
while for pullers it is contractile (fluid is pushed out from the sides and pulled in to the ends).
The shape of the elastic boundary is set by the competition between the activity of the swimmer, which tends to deform the tube and is characterised by $\tau_{\rm s}$ and $p$, 
and the elastic response of the surface, which resists such deformations and is controlled by the effective relaxation timescale $\tau_{\rm f}$. 
Note the different shape deformations for pushers and pullers. 

We characterise the effect on swimmer motility by measuring 
the ratio of the swimmer speed, $v$, to the free swimming value, $v_{\rm s}$, 
resulting from the interaction between the swimmer and the tube~[Figs.~\ref{fig:3DAll}(c) and (d)]. 
Both kinds of swimmers tend to move faster through flexible channels--characterised by small 
ratios 
of the swimming and elastic time scales, $\tau_{\rm s}$/$\tau_{\rm f}$. Even though the power consumed
by the swimmer $P$ increases with confinement, the overall swimming efficiency $\epsilon \sim v^2/P$~\cite{Note1}
increases for flexible boundaries ($\tau_{\rm s}/\tau_{\rm f} < 1$). We have found qualitatively similar results
for boundaries composed of sparse and uncoupled sets of filaments (not shown) and for single filament pairs [Fig.~\ref{fig:SnapshotsElastic}]. 

To explain these results we argue that the interaction between the swimmer and the tube can be understood in terms of {\it shift} and {\it tilt} deformations to the boundary. 
We show that shift deformations (towards the swimmer) always increase the swimmer speed
by an amount that is proportional to the self-swimming speed and independent of the strength 
or direction of the velocity field created by the swimmer. 
Conversely, tilt deformations, corresponding to a local inclination of the boundary, couple to the particular swimming 
stroke and can result in enhanced or reduced motility depending on the swimming pattern.  
Using these results we argue that the hydrodynamic coupling between the swimmer and a flexible boundary
 {\it always} leads to channel deformations which favour the passage of the swimmer.  Conversely, 
 we find that the power consumed depends on the average distance to the walls, increasing with wall proximity, and 
 not on their inclination. As a result, the swimmer speed can increase 
 due to the local deformation of the walls with the power remaining relatively constant, leading to 
 a larger swimming efficiency~[Figs.~\ref{fig:3DAll}(c) and (d)]. 
 
\begin{figure}[t!]
\psfrag{l}[c][c][0.8]{$l$}
\psfrag{h0}[l][l][0.8]{$h_0$}
\psfrag{v1}[r][r][0.8]{$v_{\rm p}$}
\psfrag{v2}[c][c][0.8]{$v_{\rm a}$}
\psfrag{v3}[c][c][0.8]{$v_{\rm s}$}
\psfrag{ha}[l][l][0.8]{$h_{\rm a}$}
\psfrag{hp}[l][l][0.8]{$h_{\rm p}$}
\psfrag{d1}[c][c][1]{$\Delta_1$}
\psfrag{d2}[c][c][1][90]{$\Delta_2$}
\includegraphics[width=0.44\textwidth]{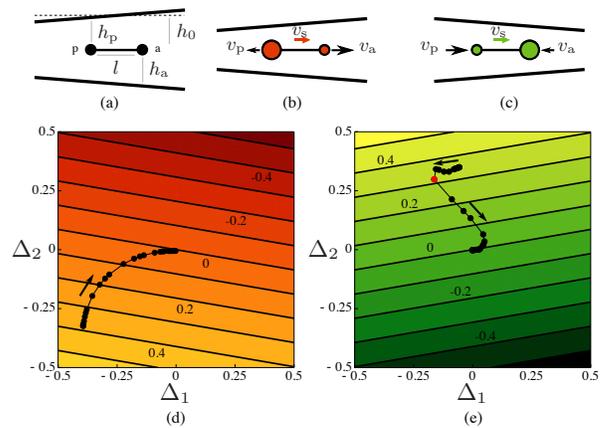}
\caption{\label{fig:Contours} (Colour online) Swimming in confinement. (a) Dipolar swimmer composed of two spheres, a and p, in the middle of
planar, inclined walls. (b) Extensile swimmers push on the posterior end of their bodies (black arrows), and experience
a speed-up when in diverging channels, light red arrow. (c) Conversely, contractile swimmers pull on their anterior end during the power step 
of their stroke (black arrows), and their speed is increased in converging channels (light green arrow). (d) and (e): Contour plots of the normalised contribution to the speed of 
pushers and pullers due to inclined boundaries, $v_{\rm i}/v_{\rm s}$, as a function of the average displacement, $\Delta_1$, and net amplitude, 
$\Delta_2$, of the channel.  The dotted paths correspond to the variation of $\Delta_1$ and $\Delta_2$ along the top curves in Figs.~\ref{fig:SnapshotsElastic}(c) and~\ref{fig:SnapshotsElastic}(d).}
\end{figure}

{\it Model swimmer:--} 
Locomotion at low Reynolds number relies
on swimming strokes that are non-reciprocal in time. 
Many microorganisms achieve this by deforming their shape in such a manner that the local drag coefficient
varies along their bodies. 
We coarse grain such a feature by considering a model swimmer made up of anterior and
posterior spheres, a and p respectively, joined by a link of prescribed but variable length 
$l$~\cite{Avron-NJPhys-2005}~[see Fig.~\ref{fig:Contours}(a)]. 
The local drag on each sphere follows from the Stokes Law, ${\bf F}_i = \xi_i {\bf v}_i,$ where 
${\bf v}_i$ is the velocity of the $i$-th sphere and $\xi_i$ its friction coefficient. 
The swimmer is subject to the force-free condition, ${\bf F}_{\rm p} + {\bf F}_{\rm a} = 0$, and to the
kinematic constraint dictated by the swimming stroke, $v_{\rm a} -  v_{\rm p} = \dot l $,
where the dot implies differentiation with respect to time, $t$.
As a consequence, 
the instantaneous speed of the swimmer obeys
$ v \equiv (v_{\rm p} + v_{\rm a})/2 = \dot l (\xi_{\rm p} - \xi_{\rm a})/2(\xi_{\rm p} + \xi_{\rm a}).$

For free, unbounded, swimmers locomotion can only be achieved if the friction 
coefficients vary in time, introducing a non-reciprocal deformation of the body of the swimmer. 
This can be readily verified by considering the Stokes drag coefficient $\xi_i = 6\pi \eta a_i$, where $\eta$ is the
viscosity of the fluid and $a_i$ is the instantaneous radius of the $i$-th sphere.  
Without loss of generality, we can set $l=l_0 + \delta l \sin(\omega t)$,    $a_{\rm p}=a_0 + \delta a \sin(\omega t + \Delta \psi + \delta \psi)$,
and $a_{\rm a}=a_0 + \delta a \sin(\omega t + \Delta \psi - \delta \psi)$ to obtain the average speed of the swimmer over one 
stroke, $ v_{\rm s} \equiv \int _0^{2\pi/\omega} {d} t v/(2\pi/\omega) = \delta l \omega \delta a \sin \delta 
\psi \cos \Delta \psi / 4 a_0 + {\cal O} (\delta a/a_0)^2$~\cite{Dunkel-SoftMatter-2010}.

The instantaneous forces exerted by the 
swimmer on the fluid can be averaged over a stroke to obtain the net flow induced by the swimmer. 
Due to the force-free condition the far-field velocity field is dipolar, $\langle {\bf u  }\rangle = {p}(3\cos^2 \phi -1){\bf \hat r}/{8\pi \eta r^2}$, 
where ${\bf r} \equiv r {\bf \hat r}$ is the displacement vector from the position of the centre of mass of the swimmer ({also stroke-averaged})~\cite{Alexander-JPhysCondMat-2009} 
to a point in the fluid, and $\phi$ measures the angle from the axis of the swimmer to $\bf r $ in the counter-clockwise direction. 
The dipole strength $p = 3 \pi \eta  \delta l \omega \delta a l_0 \cos \delta \psi \sin  \Delta \psi/2$ sets
the typical magnitude of the field, and its sign, controlled by the phase differences, 
$\Delta \psi$ and $\delta \psi$, determines whether the flow pattern is extensile ($p>0$) 
or contractile ($p < 0$).
{\it Rigid boundaries:--} 
We first focus on the interaction between the model swimmer and a bounding solid surface.  
We consider the situation where the swimmer and the surface
are aligned along a single axis of revolution, as shown in Fig.~\ref{fig:Contours}(a). For such configurations the drag 
coefficient increases, and can be written as
$\xi_i = 6 \pi \eta a_i / (1 + \zeta [a_i/h_i]),$ where $\zeta(x) < 0 $ is a correction term accounting
for the additional friction offered by the solid.  In general, $\zeta$ is a decreasing 
function of $a_i/h_i$, and its functional dependence is specific to the 
geometry of the surface.
The  interaction with the solid will lead to a net contribution to the swimming speed,  $v_{\rm w}.$ 
For planar walls ($h_i = h_0$) and prescribed deformations, the swimmer `grips' the surface, 
exploiting the higher resistance offered by the walls. {Expanding $v_{\rm w}$ 
for small swimmer deformations we find}
\begin{equation}
\label{eq:vw}
v_{\rm w } = - v_{\rm s} \left ( \frac{a_0}{h_0}\right) \frac{\zeta_0'}{1+\zeta_0},
\end{equation}
where $\zeta_0 \equiv \zeta [a_0/h_0]$, {etc}. The scaling of Eq.~(\ref{eq:vw})
indicates that the swimmer always moves faster in parallel confinement ($\zeta'_0 < 0$), by a factor controlled by the
proximity to the wall and its hydrodynamic resistance.
A similar speed-up effect has been obtained for a waving sheet swimming between 
flat walls~\cite{Katz-JFM-1974}. 

Note that Eq.~(\ref{eq:vw}) predicts that both pushers and pullers increase their speed by the same amount in parallel confining geometries. 
This is because the $h_i$ do not vary along the swimmer trajectory.
This, however, changes if the walls are inclined relative to the swimmer path. As illustration,  consider small wall inclinations
such that $h_i = h_0 + \delta_i$. { This leads} to an additional contribution
to the speed of the swimmer. { Expanding in powers of the $\delta_i$ this contribution reads}
\begin{equation}
\label{eq:vi}
v_{\rm i } =  v_{\rm s} \left ( \frac{a_0}{h_0}\right)  \left(\Delta_1 + \Delta_2 \frac{\tan \Delta \psi}{\tan \delta \psi}\right){\rm Z}_0 + { \cal O} (\Delta_i\Delta_j).
\end{equation}
The first term, which reflects the net displacement of the position of the surface from the reference value, $h_0$,  is controlled by
the {\it shift},
$\Delta_1 \equiv  (\delta_{\rm p} + \delta_{\rm a})/2h_0$. The second term results from the local inclination of the surface, 
measured by the {\it tilt}, $\Delta_2 \equiv  (\delta_{\rm p} - \delta_{\rm a})/2h_0$. Both terms scale with the function
${\rm Z_0} \equiv {\rm Z} [a_0/h_0]< 0$, which encodes the strength of the resistance provided by the wall through
${\rm Z} \equiv \left[ \zeta' - (a_0/h_0)\left(\zeta'^2/(1+ \zeta)-\zeta''\right)\right]/(1+\zeta).$

While the first term in Eq.~(\ref{eq:vi}) can be absorbed as a simple offset of $h_0$ in Eq.~(\ref{eq:vw}), the $\Delta_2$-term 
introduces an interplay between the local structure of the wall and the details of the swimming stroke, as indicated by the phase 
dependence.
Both kinds of swimmers achieve propulsion by pushing on their posterior end
during the expansion of the link ($a_{\rm p} > a_{\rm a}$), and by pulling on their anterior end 
during contraction ($a_{\rm a} > a_{\rm p}$). 
For pushers the `power step' corresponds to the
extension step, where the spheres have relatively large radii~[Fig.~\ref{fig:Contours}(b)]. Diverging constrictions, 
$\Delta_2  < 0$,  amplify the strength of this step, leading to an additional speed-up. For converging 
constrictions, however, the power step is weakened by an amount controlled by $\Delta_{2}$, which
can lead to a net slow-down. Similarly,  contractile swimmers generally swim faster in converging 
constrictions~[Fig.~\ref{fig:Contours}(c)], 
relative to the planar wall reference configuration.

The interplay between wall proximity and wall inclination can lead to both speed-ups and slow-downs for the swimmer~[see contour plots in Figs.~\ref{fig:Contours}(d) and~\ref{fig:Contours}(e)].  
Note that $v_{\rm i}$ can be of the same order of magnitude as $v_{\rm s}$.
To obtain these plots we have set ${\rm Z}_0$ according to Fax\'en's correction 
to Stokes Law~\cite{Faxen-Ann-1922}, whereby $\zeta(x) = A x + B x^3 + Cx^5,$ with $A=-1.004$, $B=0.418$ and $C=-0.169.$ This particular
friction law is valid for a pair parallel plates, relevant for swimmers confined in microfluidic chambers, and is expected to hold for 
our treatment in the limit of gently, locally, inclined surfaces.

{The additional drag induced by the walls causes a net increase in the average rate of energy consumed
by the swimmer over a cycle, $P = \int_0^{2\pi/\omega} d t (F_{\rm a}v_{\rm a}+F_{\rm p }v_{\rm p})/(2\pi/\omega) $. 
For parallel boundaries we find $P_{\rm w} \approx P_{\rm s}/(1+\zeta_0)$, larger than the free-swimming value 
$P_{\rm s}$ by a factor increasing for stronger confinement. 
However, for tapered channels the leading order contributions to the total power consumption depend only on the boundary shift, $\Delta_1$,
{\it i.e.},  
\begin{equation}
P = P_{\rm w}\left(1+\Delta_1\left(\frac{a_0}{h_0}\right)\frac{\zeta'_0}{1+\zeta_0}\right) + {\cal O}({\Delta_i\Delta_j}),
\end{equation}
and
are unaffected by the inclination of the walls.  As a consequence, the speed of the swimmer 
can vary due to the kinematic coupling between the swimming stroke and the orientation
of the boundaries at constant power consumption.} 

{\it Elastic boundaries:--} 
{We now test the applicability of the {\it shift-tilt} theory to understand the enhanced motility of
swimmers observed in our simulations of flexible channels.
Qualitatively, pushers deform
the boundary to create a locally diverging channel, while contractile swimmers induce a local converging geometry~[see Figs.~\ref{fig:3DAll}(a) and (b)]. These configurations
lead to the expected speed-up, according to our arguments. 
\begin{figure}
\centering
\psfrag{v}[c][c][1][90]{$v_{\rm f} /{ v_{\rm s}}$}
\psfrag{W}[c][c][1][0]{$x/l$}
\psfrag{Y}[c][c][1][180]{$ v_{\rm f} / v_{\rm s}$}
\psfrag{R}[c][c][1][180]{$ y/l$}
\psfrag{X}[c][c]{$\tau_{\rm s}/\tau_{\rm f}$}
\psfrag{p1}[r][l][0.5]{$\delta_{\rm 0}/h_{\rm 0} = 1.04$}
\psfrag{p2}[r][l][0.5]{$\delta_{\rm 0}/h_{\rm 0} = 0.51$}
\psfrag{p3}[r][l][0.5]{$\delta_{\rm 0}/h_{\rm 0} = 0.18$}
\psfrag{y}[c][c][1][180]{$y/h_0$}
\psfrag{x}[t][t]{$x/h_0$}
\psfrag{d1}[l][l][0.8]{$\delta_{\rm p} / h_0 $}
\psfrag{d2}[l][l][0.8]{$\delta_{\rm a} / h_0 $}
\psfrag{L}[c][c][1]{$\lambda/h_0 $}
\psfrag{v1}[l][l][0.8]{$v_{\rm s} $}
\includegraphics[width=0.44\textwidth]{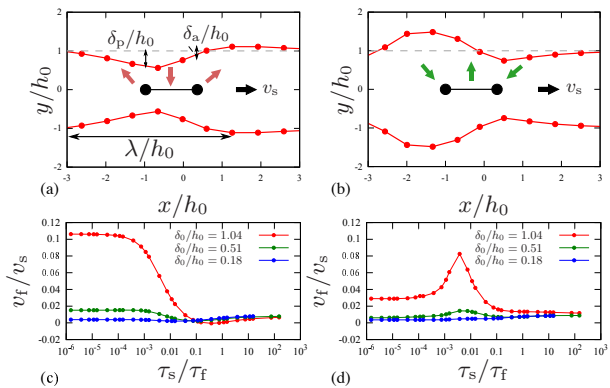}
\caption{\label{fig:SnapshotsElastic} (Colour online) Typical deformations of an elastic boundary caused by (a) pushers and (b) pullers.  Due to the
flow-field pattern, shown as light coloured arrows, the interface is deformed to a locally diverging or converging shape for pushers and pullers, respectively. 
The local deformation
of the filaments is quantified by interpolating $\delta_{\rm p}$ and $\delta_{\rm a}$ at the $x-$coordinate of the respective beads.
(c) and (d) Corresponding change in speed, $v_{\rm f} = v - v_{\rm s}$, relative to the free-swimming speed. The amplitude of the response is controlled
by the shape of the boundary, which is set by the ratio of swimmer and filament timescales, $\tau_{\rm s}/\tau_{\rm f},$ and by the swimmer dipolar
strength, $p$ $(\delta_0\sim p).$}
\end{figure}
Two main effects set the deformation of the filaments, and in turn the back flow $v_{\rm f}\equiv v-v_{\rm s}$. On the one hand, the filament is deformed by the 
velocity field of the passing swimmer, of typical magnitude $v_0 \sim p /\eta h^2_0$, over the swimmer self-propagation timescale $\tau_{\rm s} = l_0/v_{\rm s}$.
On the other, the elastic resistance of the filament can be quantified by the bending relaxation timescale 
$\tau_{\rm f} \sim \eta a_{\rm f} \lambda^3 / G,$ 
where $\lambda \sim h_0$ is the wavelength of the perturbation,  $G\approx k_{\rm b} r_{\rm eq} $ is the bending modulus
and $r_{\rm eq}$ is the rest length between points in the filament. 
The growth rate of perturbations to the boundary thus obeys, $\dot \delta \approx v_0 - \delta/\tau_{\rm f},$ which 
suggests a scaling 
$\delta \sim \delta_0 \left[1- \exp (-\tau_{\rm s}/\tau_{\rm f})\right]/(\tau_{\rm s}/\tau_{\rm f}),$
where $\delta_0 \equiv v_0 \tau_{\rm s}$ is the typical amplitude for freely-deformable chains. 
To explore the interplay between these effects in detail, we ran simulations considering a pair of initially parallel filaments [Figs.~\ref{fig:SnapshotsElastic}(a) and (b)] 
and explored a wide range in $\tau_{\rm s}/ \tau_{\rm f}$. This simple configuration only reduces the magnitude of 
the back flow with respect to the many-filament simulations, and does not alter results qualitatively. 
We focus on the dependence of $v_{\rm f}$ on $\tau_{\rm s}/\tau_{\rm f}$ for pushers and pullers of identical 
free-swimming speed $v_{\rm s}$ but with different dipole strength, $p$~[Figs.~\ref{fig:SnapshotsElastic}(c) and (d)] . As suggested by the simple scaling argument, curves tend to reach 
the rigid-boundary limit at $\tau_{\rm s}/\tau_{\rm f} \approx 1$, when the rigidity of the filaments suppresses deformations. Additionally,
the amplitude of the curves scales with $\delta_0$, as expected for swimmers with stronger dipole moments that can induce larger deformations
to the boundary. 

The shape of the curves shown in Fig.~\ref{fig:SnapshotsElastic} is set by the subtle interplay between the local inclination 
and shift induced by the swimmer on the elastic boundary. To illustrate this point, we measured the local amplitudes $\delta_{\rm p}$
and $\delta_{\rm a}$ [see Fig.~\ref{fig:SnapshotsElastic}(a)] for the top curve in Figs.~\ref{fig:SnapshotsElastic}(c) and~(d). We
then calculated the corresponding $\Delta_1$ and $\Delta_2$ values, which we depict as paths in $(\Delta_1,\Delta_2)-$space
in Figs.~\ref{fig:Contours}(d) and~\ref{fig:Contours}(e). 
Extensile swimmers tend to pull on the boundary, shifting it to closer positions and to higher inclinations simultaneously. Both effects 
decrease with increasing filament rigidity, leading to a trajectory in $(\Delta_1,\Delta_2)-$space that runs monotonically from negative values
of both parameters to the origin, and is reflected in the smooth decrease of $v_{\rm f}$ with $\tau_{\rm s}/\tau_{\rm f}$ 
shown in Fig.~\ref{fig:SnapshotsElastic}(c). 
Contractile swimmers, on the contrary, strongly deform the boundary, creating high local inclinations, but only shift its position 
weakly.  For intermediate rigidity of the filament, while the inclinations die out, the boundary is shifted to a closer position 
to the swimmer. This contributes to speeding it up. However, for higher rigidity the shift decays and this effect vanishes. This competition 
leads to a turning trajectory in $(\Delta_1,\Delta_2)$-space which shows a good correlation with the non-monotonic behaviour observed 
for $v_{\rm f}$$-$the maximum of the top curve of Fig.~\ref{fig:SnapshotsElastic}(d) corresponds to the 
right-most point of the path in $(\Delta_1,\Delta_2)$-space (red dot) in Fig.~\ref{fig:Contours}(e). 

We have described the hydrodynamic coupling between the self-propulsion of microscopic swimmers and the resistance offered 
by elastic confining surfaces.  
Both pushers and pullers deform the surfaces in such a way that enhances their motility. 
A similar microscopic mechanism is likely to underlie the enhanced motility 
observed in viscoelastic fluids~\cite{Lauga-RepProgPhys-2009,Gagnon-EPL-2013}.
We have also shown that locally inclined constrictions can enhance or hinder
the motility of swimmers depending on their swimming stroke. 
{This asymmetry has potential as a means of separation or selective permeation of microorganisms.  
We hope that our research will motivate experiments to explore these possibilities.}

We thank Henry Shum and Mitya Pushkin for fruitful discussions. 
R.L.-A. acknowledges support from Marie Curie Actions (FP7- PEOPLE-IEF-2010 no. 273406),  and JMY from the ERC Advanced Grant (MiCE).\\


\end{document}